\documentclass[aps,prc,superscriptaddress,showpacs,manuscript]{revtex4}
\usepackage{amsmath,bm}%
\usepackage{graphicx}%
\begin{document}
\title{A novel approach to Isoscaling: the role of the order parameter 
$m=\frac{N-Z}{A}$}
\author{M. Huang}
\affiliation{Cyclotron Institute, Texas A$\&$M University, 
College Station, Texas 77843}
\affiliation{Institute of Modern Physics, Chinese Academy of Sciences, 
Lanzhou, 730000,China.}
\affiliation{Graduate University of Chinese Academy of Sciences,
 Beijing, 100049, China.}
\author{Z. Chen}
\affiliation{Cyclotron Institute, Texas A$\&$M University, 
College Station, Texas 77843}
\affiliation{Institute of Modern Physics, Chinese Academy of Sciences, 
Lanzhou, 730000,China.}
\author{S. Kowalski}
\affiliation{Institute of Physics, Silesia University, Katowice, Poland.}
\author{R. Wada}
\affiliation{Cyclotron Institute, Texas A$\&$M University, 
College Station, Texas 77843}
\author{T. Keutgen}
\affiliation{FNRS and IPN, Universit\'e Catholique de Louvain, B-1348 
Louvain-Neuve, Belgium.}
\author{K. Hagel}
\affiliation{Cyclotron Institute, Texas A$\&$M University, 
College Station, Texas 77843}
\author{J. Wang}
\affiliation{Institute of Modern Physics, Chinese Academy of Sciences, 
Lanzhou, 730000,China.}
\author{L. Qin}
\affiliation{Cyclotron Institute, Texas A$\&$M University, 
College Station, Texas 77843}
\author{J.B. Natowitz}
\affiliation{Cyclotron Institute, Texas A$\&$M University, 
College Station, Texas 77843}
\author{T. Materna}
\affiliation{Cyclotron Institute, Texas A$\&$M University, 
College Station, Texas 77843}
\author{P.K. Sahu}
\affiliation{Cyclotron Institute, Texas A$\&$M University, 
College Station, Texas 77843}
\author{M.Barbui}
\affiliation{Cyclotron Institute, Texas A$\&$M University, 
College Station, Texas 77843}
\author{C.Bottosso}
\affiliation{Cyclotron Institute, Texas A$\&$M University, 
College Station, Texas 77843}
\author{M.R.D.Rodrigues}
\affiliation{Cyclotron Institute, Texas A$\&$M University, 
College Station, Texas 77843}
\author{A. Bonasera}
\email[E-mail at:]{bonasera@lns.infn.it}
\affiliation{Cyclotron Institute, Texas A$\&$M University, 
College Station, Texas 77843}
\affiliation{Laboratori Nazionali del Sud, INFN,via Santa Sofia, 
62, 95123 Catania, Italy}


\date{\today}
 
\begin{abstract}
Isoscaling is derived within a recently proposed modified Fisher model 
where the free energy near the critical point is 
described by the Landau $O(m^6)$ theory. In this model $m=\frac{N-Z}{A}$ 
is the order parameter, a consequence of (one of) the symmetries
of the nuclear Hamiltonian.  Within this framework we show that isoscaling 
depends mainly on  this order parameter through the 'external (conjugate) 
field' H. The external field is just given by the difference in chemical 
potentials of the neutrons and protons of the two sources. To distinguish 
from previously employed isoscaling relationships, 
this approach is dubbed: $m-scaling$. We discuss  the relationship  between  this framework and the standard isoscaling formalism  and point out some substantial differences in interpretation of experimental results which might result. These should be investigated further 
both theoretically and experimentally. 
\end{abstract}
\pacs{21.65.Ef, 24.10.-i, 24.10.Pa,25.70.Gh, 25.70.Pq}
\maketitle

\section *{ INTRODUCTION}

In near Fermi energy heavy ion collisions fragments are copiously 
produced. 
The mass distributions of these fragments often exhibit a power law 
behavior~\cite{Minich82,bon00}. The isotopic distribution of these 
fragments is governed by the free energy at the density and temperature 
of the emitting system.  We have recently  
discussed the isotope production  in terms of the Modified Fisher 
Model~\cite{Bonasera08}. The experimental results exhibit a dependence on 
an order parameter $m=\frac{N-Z}{A}$.  This analysis and 
the often reported observation of isoscaling for products of two similar
reactions with different neutron to proton ratios, N/Z~\cite{Xu00,Tsang01,
Botvina02,Ono03,Bonasera08}  make it clear that terms in the free energy 
which are sensitive to the difference in neutron and 
proton concentrations are very important in the fragment formation process.
Indeed, isoscaling analyses based on the comparison of isotope yields from 
excited systems of similar temperatures and Z have been employed to 
obtain information on the symmetry energy and its density 
dependence~\cite{Xu00,Tsang01,Botvina02,Ono03,chen}.
The ratio of the isotope yields,
R$_{12}$, between two similar reaction systems with  different $N /A$ 
ratios can be expressed by the following isoscaling 
relation~\cite{Xu00,Tsang01}:
\begin{equation}
R_{12}(N,Z)=Cexp(\alpha N + \beta Z),
\label{eq:isoscaling}
\end{equation}
where the isoscaling parameters,  $\alpha = (\mu_{n}^{1} - \mu_{n}^{2})/T$
and $\beta = (\mu_{p}^1 - \mu_{p}^2)/T$, represent the differences of the 
neutron ( or proton) chemical potentials between systems 1 and 2, 
divided by the temperature. C is a constant.
 
In terms of a modified Fisher model description the experimental yield of 
an isotope with N neutrons and Z protons
can be written as~\cite{Minich82,Bonasera08,bon00}:
\begin{eqnarray}
Y(N,Z)=Y_0A^{-\tau}exp\{-[G(N,Z) \nonumber \\
-\mu_nN-\mu_pZ]/T\},
\label{eq:yield}
\end{eqnarray}
where Y$_{0}$ is a constant, G(N,Z) is the nuclear free energy at the time 
of the fragment formation, $\mu_{n}$ and $\mu_{p}$ are the neutron and 
proton  chemical potentials, and T is the temperature of the emitting 
source.
The factor, $A^{-\tau}$,
originates from the entropy of the fragment~\cite{Minich82}. 
Notice that in the ratio of yields employed in an isoscaling analysis  
the power law term from Eq.(\ref{eq:yield}) will cancel out.

In a grand canonical treatment the relationship between the isoscaling  
may be expressed as~\cite{Botvina02,Ono03}:
\begin{equation}
\alpha(Z) = 4C_{sym}\Delta (Z_s/{A_s})^{2}/T,
\label{eq:alpha}
\end{equation}
where $\Delta(Z_s/{A_s})^2 = {(Z_s/{A_s})^{2}}_{1} 
- {(Z_s/{A_s})^{2}}_{2}$, $C_{sym}$ is the symmetry free energy 
and T is the temperature. 
The symmetry free energy is presumed to be density dependent. 
In a similar fashion $\beta(N)$ can be expressed as :
\begin{equation}
\beta(N) = 4C_{sym}\Delta (N_s/{A_s})^{2}/T,
\label{eq:beta}
\end{equation}
where $\Delta(N_s/{A_s})^2 = {(N_s/{A_s})^{2}}_{1} 
- {(N_s/{A_s})^{2}}_{2}$.
This suggests that for two systems at temperature T  in which the symmetry 
energy is the dominant factor in determining the yield ratios of the 
emitted fragments, the ratio $\beta(N)/\alpha(Z)$ can be expressed as  
\begin{equation}
\eta=\frac{\beta(N)}{\alpha(Z)} = -\frac{(N_s/A_s)^2_1 - (N_s/A_s)^2_2}{(Z_s/A_s)^2_1 - 
(Z_s/A_s)^2_2},
\label{eq:kai}
 \end{equation}
This ratio will approach  -1 when systems of very similar N/Z are 
considered but in general can be quite different from -1.

It is well known that nucleon forces are isospin invariant, 
because of this we expect that, in the absence of spontaneous symmetry 
breaking, $(\mu_{n}^{1} - \mu_{n}^{2}) = (\mu_{p}^1 - \mu_{p}^2)$ 
and hence 
\begin{equation}
\alpha=-\beta.
\label{eq:invarian}
\end{equation}
In a manner similar to the case for mirror nuclei, at low excitation 
energies we can expect this invariance to be broken by Coulomb energy 
contributions. However in fragmentation reactions occurring near 
the critical point of the liquid-gas phase change the nuclear symmetry 
should be restored because of the invariance of the nuclear Hamiltonian 
when $m\rightarrow -m$. 
In a separate paper discussing the analysis of the same data set used 
in this paper, a clear fragment Z (or N) dependence of $\alpha(Z)$ (or $\beta(N)$) 
has been reported~\cite{chen,huang10}.
>From the detailed comparisons to the dynamical 
model (AMD) calculation followed by the statistical decay code (Gemini), 
this Z dependence is attributed to the statistical secondary 
decay process of the excited fragments after they are 
formed at freezeout of the emitting source. A significant 
modification of the isoscaling parameters is also suggested 
during the cooling process. Similar results have been also reported 
in ref.~\cite{Botvina02}. In that work it is also concluded that secondary 
decay effects play a significant role in determining the observed ratio. 
Their conclusion is based upon the application of an SMM model 
description to the experimental data within a statistical model 
description of fragmentation. In this paper we suggest that there 
may be an essential relatioship between the scaling parameter 
and N/Z of the emitting system, which is related to the restoration 
of symmetry near the critical point of the emitting source  and which is 
sustained in the experimental observables during the 
cooling process and  experimentally manifested in the isoscaling 
parameters.

\section{Experiments and Analysis}

Using high resolution detector telescopes with excellent isotope 
identification capabilities, we recently studied a number of heavy 
ion reactions to determine relative yields for production of a wide 
range of isotopes~\cite{chen,huang10}.
The experiment was performed at the K-500 superconducting cyclotron 
facility at Texas A$\&$M University. 40 A MeV $^{64}$Zn,$^{70}$Zn and 
$^{64}$Ni beams irradiated $^{58}$Ni,$^{64}$Ni, $^{112}$Sn,$^{124}$Sn,
$^{197}$Au, and $^{232}$Th targets. 
Intermediate mass fragments (IMFs) were detected by a detector telescope 
placed at 20 degrees relative to the beam direction. The telescope 
consisted of four Si detectors. Each Si detector had an area of  5cm 
$\times$ 5cm. 
The thicknesses are 129, 300, 1000 and 1000 $\mu$m. 
Using the 
$\Delta E-E$ technique we were typically able to identify 6-8 isotopes 
for a given Z up to Z=18 with  energy thresholds of 4-10 A MeV.
More details of the analysis are contained in ref.~\cite{chen,huang10}.

Isoscaling analyses were carried out for all possible combinations of 
these reactions. Eighteen different reactions are considered here and therefore 
more than 150 combinations are studied.
Data for each atomic number were independently fit to extract the 
isoscaling parameter $\alpha(Z)$. $\beta(N)$ values were also extracted 
for each neutron number. For some systems the extracted $\alpha(Z)$ parameter shows a steady decrease as Z increases. The $\beta(N)$ parameter generally showsa much smaller variation with increasing N, and has the opposite sign. A clear correlation between them, i.e. $\alpha(Z) \sim -\beta(N)$ for the equivalent number of nucleons, $N = Z$, has also been observed as suggested in the introduction(see Eq.(\ref{eq:invarian})). In Fig.1, the extracted isoscaling parameters for the case of Z = N = 7 are shown as a typical example. 
Similar correlations are also observed for other selections of Z and N values if Z = N.
\begin{figure}[ht]
\includegraphics[width=3.6in,clip]
{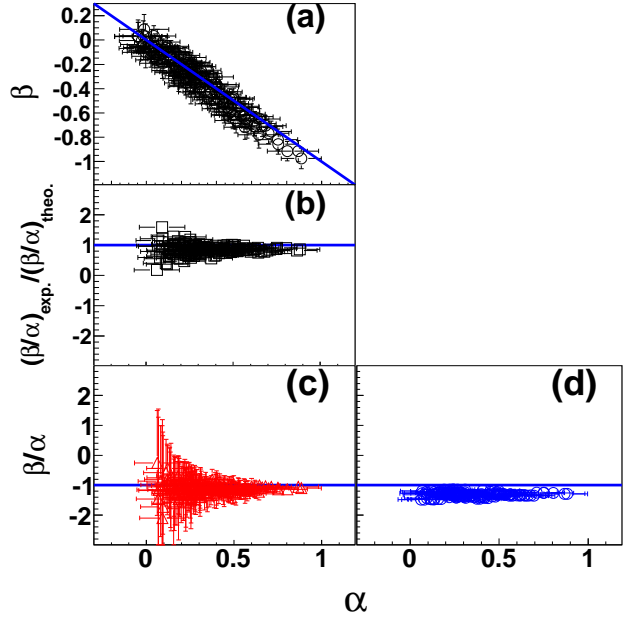}
\caption{\label{fig:fig_1}(a) $\beta(N)$ vs $\alpha(Z)$ is plotted for N=Z=7. Line indicates
the locus $\beta(N) =- \alpha(Z)$;
(b)ratio of $\beta(N)/\alpha(Z)$ data to theoretical Eq.(\ref{eq:kai}) (open squares); 
(c)$\beta(N)/\alpha(Z)$ vs. $\alpha(Z)$ from data (open triangles); 
(d)the analytical prediction Eq.(\ref{eq:kai}) (open circles) to compare with data.  }
\end{figure}
    

In part a of Figure 1 we have plotted $\beta(N)$ vs $\alpha(Z)$. 
As seen in the figure, the relation $\alpha(Z) \sim -\beta(N)$ is observed 
for $\alpha(Z) \le 0.5$
and may deviate slightly at the larger $\alpha(Z)$ values.  
Those larger values are associated with the largest N/A values 
for the compound system. In the bottom part of the figure, these values on the left are 
compared to predictions of Eq.(\ref{eq:kai}) on the right with the assumptions that Z/A 
is that of the compound system. 
We note that the experimental values tend to be significantly closer 
to -1 than the calculated values. 
Except at the low experimental values of $\alpha(Z)$ where the scatter 
is significant,  the experimental values for  $\beta(N)/\alpha(Z)$ are about 
$20\%$ lower in the absolute value than the model values as indicated by the ratio of these two 
quantities also  plotted in the middle part of the figure.
In order to see the 
system dependence of $\alpha(Z)$ and $\beta(N)$ values, these values are 
plotted for separate groups of fissility values in Fig.2.
The  fissility is defined as $X=\frac{Z_s^2}{A_s}$, where
$Z_{s}$ and $A_{s}$ are the charges and masses of the
source which we assume to be the compound nucleus for simplicity.  
We can define a combined fissility parameter between reactions (1) 
and (2) as $\Sigma X=X_1+X_2$.
Larger (absolute) values of $\alpha(Z)$ and $\beta(N)$ correspond to 
large values of the $\Sigma X$ parameter. 
In the figure $\alpha(Z)$ and $\beta(N)$ values are separately plotted for four different 
ranges of fissility group for the same data set used in Fig.1. 
We see no systematic correlation with the fissility parameters in the 
deviation from $\alpha(Z) = -\beta(N)$,
which might be suggestive of the fact that the Coulomb force is not 
so effective for breaking the invoked invariance of the Nuclear Hamiltonian. 
It would be very interesting to see if Coulomb effects become more 
important for heavy colliding nuclei such as $U+U$.
One should note that a similar result is observed for  
$\alpha(Z)$ and $\beta(N)$ in other IMFs, when Z and N are the same.
One can also use  $\alpha$ and $\beta$ values averaged over a range of
atomic (or neutron) number, though in this case 
the averaged numbers depend on the somewhat range selected.    
\begin{figure}[ht]
\includegraphics[width=3.6in]
{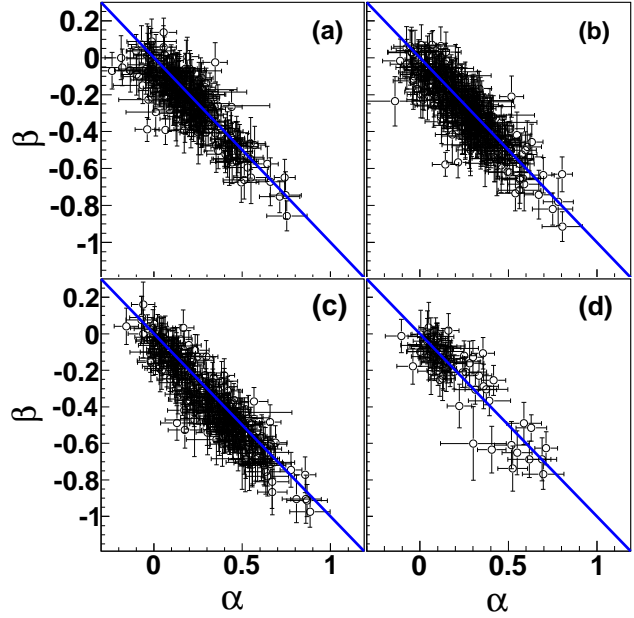}
\caption{\label{fig:fig_2}$\beta(N)$ vs $\alpha(Z)$ with $6 \le Z \le 13$ for different ranges of the 
fissility parameter.  $\Sigma X\le60.9$ (top left), 
$60.9<\Sigma X\le72.2$ (top right), 
$72.2<\Sigma X\le83.4$ (bottom left)
and $83.4<\Sigma X\le94.7$ (bottom right). }
\end{figure}


While the trend in Figs.\ref{fig:fig_1} and \ref{fig:fig_2} is interesting, it is important to 
note that when the neutron and proton concentrations of the initial 
excited source are different, two well established trends can act to 
shift the balance toward symmetric matter and hence to bring the 
absolute values of the observed $\alpha(Z)$ and $\beta(N)$ parameters closer 
together. The first is the distillation effect in which early emission of 
particles favors neutron emission over proton emission~\cite{serot,maria}.
As a result of this early emission the fragmenting system will tend 
to have a higher symmetry than the initial system. The second is 
secondary decay of initially excited fragments~\cite{Marie98,Hudan03}
which favors a shift toward the evaporation attractor line~\cite{Charity88}.
 Thus even if the comparison of primary fragment yields
would lead to a significant difference in the two isoscaling 
parameters the subsequent decay can reduce this difference. 



\section *{$m$-SCALING}

Pursuing the question of phase transitions, we note that  
we have previously discussed some of the present yield data within 
the Landau free energy  description~\cite{Bonasera08}. In this 
approach the ratio of the free energy (per particle) to the temperature 
is written in terms of an expansion:
\begin{equation}
  \label{eq:order}
 \frac{F}{T}=\frac{1}{2}a m^2+\frac{1}{4}b m^4 
+\frac{1}{6}c m^6-m\frac{H}{T},
 \end{equation}
where $m$ is  the order parameter, $H$ is its conjugate variable and 
$a-c$ are fitting parameters. In our case  $m = (I/A)$.
Notice that the free energy that we have indicated 
with F includes the chemical potential of neutrons and protons i.e. 
$AF(m,T)=[G(N,Z)-\mu_nN-\mu_pZ]$ (compare to Eq.(\ref{eq:yield})). 

We observe that the free energy is even in the exchange of 
$m \rightarrow -m$ reflecting the invariance of the nuclear
forces when exchanging N and Z.  This symmetry is violated by 
the conjugate field $H$ which arises when the source is asymmetric
in the chemical composition.  We stress that correctly m and H are 
related to each other through the relation $m=-\frac{\delta F}{\delta H}$.

An immediate consequence of the application of the Landau expression of 
Eq.(\ref{eq:order}) in the Modified Fisher Model is that it brings a 
scaling law for m=0 isotopes. Since F(m=0,T)=0, for any T,
the yield in Eq.(\ref{eq:yield}) is given as
\begin{eqnarray} {
Y(N,Z)=Y_0A^{-\tau},
}
\label{eq:yield_m0}
\end{eqnarray}
for all reactions. 
\begin{figure}[ht]
\includegraphics[width=3.6in]{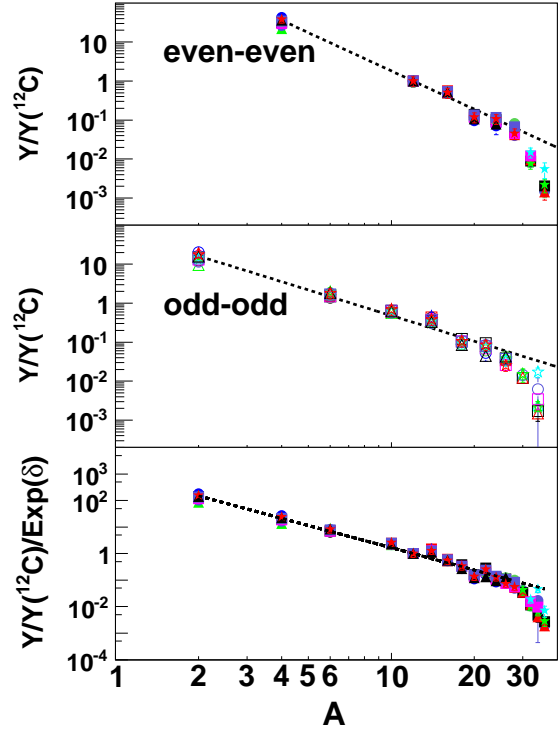}
\caption{\label{fig:fig_3}Yield ratio of m=0 isotopes vs A. The yield is 
normalized to that of $^{12}C$. Data from all 18 reaction 
systems studied in this experiment. (top) even-even 
isotopes. (middle) odd-odd isotopes are plotted. 
(bottom) pairing corrected yield, Y(N,Z)/exp($\delta$), are plotted for 
all m=0 isotopes. Lines in each figure are linear fitted ones. 
$\tau$ values are 3.3, 2.2 and 2.8 from the top to bottom. 
$a_p/T$ =2.2 is used 
in the bottom.}
\end{figure}
In Fig.\ref{fig:fig_3}, yield ratios for m=0 isotopes are separately plotted as a function of A for even-even (top) 
and odd-odd (middle) isotopes for all 18 reactions studied here. 
In order to eliminate the effect of the constant in Eq.(\ref{eq:yield}), 
which are slightly different in each reaction system, the yield is 
normalized to that of the $^{12}$C in each reaction. 
As seen in the figure, the yields from the different reactions are indeed 
scaled well with A up to A $\sim$ 30 when even-even and odd-odd isotopes are plotted 
separately. One should note that data points for a given A represent 
all 18 different reactions in the figure. The slope difference between even-even 
and odd-odd isotopes can be naturally attributed to the pairing effect. 
However a large pairing effect is expected only at a low temperature, because it is related to the shell effect. 
On the other hand the emitting sources of these isotopes are expected to be
at a high temperature. Ricciardi et al.  
have given a possible explanation for this observation~\cite{Ricciardi04,Ricciardi05}. 
According model simulations which they have performed the experimentally observed 
pairing effect is attributed to the last chance particle decay 
of the excited fragments during their cooling. This hypothesis is
also supported by our model simulations presented in a separate 
paper~\cite{huang10}. 
In order to take into account the pairing effect, data for 
even-even and odd-odd isotopes were simultaneously fitted by the following equation,
\begin{eqnarray}
Y(N,Z)=Y_{0}A^{\tau}exp(\delta/T),
\label{eq:yield_m0_delta}
\end{eqnarray}
\begin{eqnarray}
\delta(N,Z)=\left\{\begin{array}{ll} 
 a_{p}/A^{1/2 }&(\textrm{odd-odd})\\
 0 &(\textrm{even-odd})\\       
-~a_{p}/A^{1/2 }&(\textrm{even-even} ).
\end{array}\right.
\label{eq:paringterm}
\end{eqnarray}
and the parameters $\tau$ and $a_p/T$ values was extracted. Using these extracted 
parameters, the experimental yield was divided by the exponent 
in Eq.(\ref{eq:yield_m0_delta}) as factor. The results are plotted in the bottom 
of the figure for all isotopes with m=0. The extracted $\tau$ value is 2.8 
which is larger than the normal critical exponent 2.3.
This difference may reflect either  the temperature of the emitting 
source is below the critical temperature or that the secondary decay processes 
modify the value. 
 
Because of the symmetries of the free energy when we take the ratio
 between two different systems, $presumably$ at the same temperature 
$T$ and density $\rho$, all $even$ 
order terms in $m$ cancel out while the $odd$ terms remain.  
Those terms depend on the $external$ field $H/T$.  
Taking the ratio between two systems as in Eq.(\ref{eq:yield}) we easily 
obtain:
\begin{equation}
R_{12}(m)=Cexp(\Delta H/T m A),
\label{eq:mscaling}
\end{equation}
where $\Delta H/T=H_1/T-H_2/T$. We can fix the constant C by dividing 
each experimental yield by the $^{12}C$ yield following 
in ref.~\cite{Bonasera08}. The goal is to get $C\rightarrow1$ for 
reasons that will become clear below.
Comparing the latter equation with Eq.(\ref{eq:isoscaling}) we obtain: 
$\Delta H/TmA=\alpha N + \beta Z$ i.e. $\alpha=-\beta=\frac{\Delta H}{T}$.
As shown in Fig.\ref{fig:fig_1}, for the comparison for isotopes of a given Z with isotones having N equal to that Z, this relation appears to be satisfied. 
The relation is valid more in general, and in fact we could
write the chemical potentials of neutrons and protons as:

 \begin{equation}
\mu_nN+\mu_pZ= \mu A+H m A,
\label{eq:muu}
 \end{equation}
from this relation it follows that:
\begin{equation}
2H=\mu_n-\mu_p;  
 2\mu=\mu_n+\mu_p.
\label{eq:muuh}
 \end{equation}
All these relations show that if $m$ is an order parameter then 
$\alpha=-\beta = \Delta H/T$.  


\begin{figure}[ht]
\includegraphics[width=3.6in]{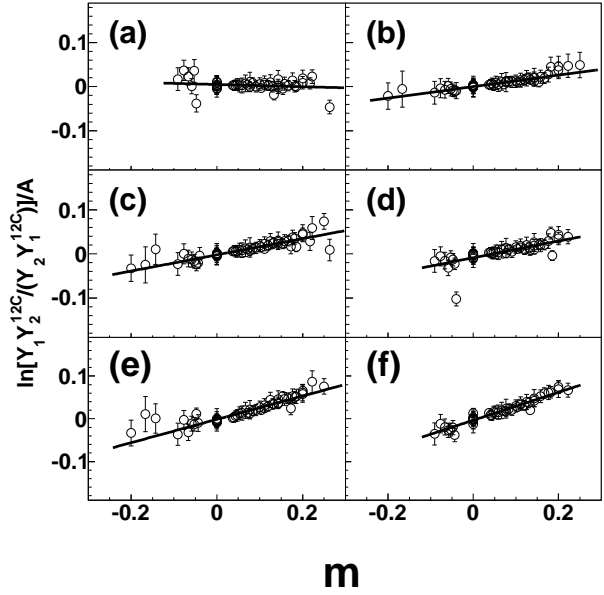}
\caption{\label{fig:fig_4}Experimental ratios vs m for isotopes 
with $6 \le Z \le 13$ 
for (a) $\frac {^{64}Ni^{232}Th}{^{70}Zn^{197}Au}$,  
(b)$\frac{^{64}Ni^{112}Sn}{^{64}Ni^{58}Ni}$,
(c)$\frac{^{64}Ni^{124}Sn}{^{64}Ni^{64}Ni}$,  
(d)$\frac{^{64}Ni^{197}Au}{^{64}Ni^{112}Sn}$,  
(e)$\frac{^{64}Ni^{124}Sn}{^{70}Zn^{58}Ni}$  and 
(f)$\frac{^{64}Ni^{124}Sn}{^{70}Zn^{112}Sn}$ respectively at
 $40 MeV/A$.
 The lines are the results of a linear fit according to Eq.(\ref{eq:mratio}).
}
\end{figure}

The external field is given by the difference of chemical potentials
between neutrons and protons of the emitting system as expected.  
>From Eq.(\ref{eq:mscaling}) we can obtain the difference between 
the free energies (or alternatively the external fields) as:
 \begin{equation}
\frac{-ln(R_{12}(m))}{A}=\Delta H/T m +constant.
\label{eq:mratio}
\end{equation}
Thus a plot of $-ln(R)/A$ versus m should give a linear relation 
whose slope is given by $\Delta H/T$. 
This linear relation is demonstrated in Figs.\ref{fig:fig_1} 
and \ref{fig:fig_2} where such a plot 
is obtained for different colliding systems for the isotopes in 
the selected range of Z. In thatgiven range $\alpha(Z)$ increases
about 50\% on average~\cite{chen,huang10}. As discussed in references~\cite{chen,huang10}, 
the observed fragment Z (or N) dependence of the isoscaling parameters is mainly established   
during the statistical cooling  of the excited fragments. 
In fact it has been demonstrated that $\alpha(Z)$ parameter extracted from 
the primary fragments of the AMD simulations shows no significant 
fragment Z dependence.
It should be noted that it is important to normalize the distribution 
( for instance to $^{12}C$ ) as we have done in order that the 
normalizing constant in front of the yield in Eq.(\ref{eq:mscaling}) 
is one. If not this will carry a $\frac{1}{A}$ term which might violate 
the scaling.
Overall the scaling is satisfied for this set of data as seen in Fig.\ref{fig:fig_4}.
Compared to 'traditional' isoscaling where a fit is 
performed for each detected charge $Z$ (or each $N$) we see that all the 
data collapse into  one curve.

We can further elucidate the role of the external field $H/T$ writing the Landau expansion and 'shifting' the order parameter by $m_s$ which is the position of the minimum of the free energy.
Such a position  depends on the neutron to proton concentration of the source~\cite{Bonasera08}.  Thus 
\begin{equation}
  \label{eq:shifted}
 \frac{F}{T}=\frac{1}{2}a (m-m_s)^2+\frac{1}{4}b (m-m_s)^4 +\frac{1}{6}c (m-m_s)^6.
 \end{equation}
Comparing to Eq.(\ref{eq:order}) we easily obtain
\begin{equation}
  \label{eq:hms0}
 \frac{H}{T}=(a + b m^2 + c m^4 )m_s 
+(\frac{1}{2}\frac{a}{m}+\frac{3}{2}b m +\frac{5}{2}c m^3)m_s^2 + O(m_s^4)...,
 \end{equation}
thus $H$ depends on the source isospin concentration though the 
parameter $a, b, c$ which are terms of the free energy.  
We stress that these terms refer to the free energy and $not$ 
to the internal symmetry energy. If b and c are of comparable magnitude to parameter a, then taking terms of a, Eq.(\ref{eq:hms0}) can be further simplified as
\begin{equation}
 \label{eq:hms}
 \frac{H}{T}=-a m_s +\frac{1}{2}a\frac{m_s^2}{m}+O(m_s^4)...,
\end{equation}

 \section{Reconciliation of the two approaches} 

\begin{figure}[ht]
\includegraphics[width=3.6in]{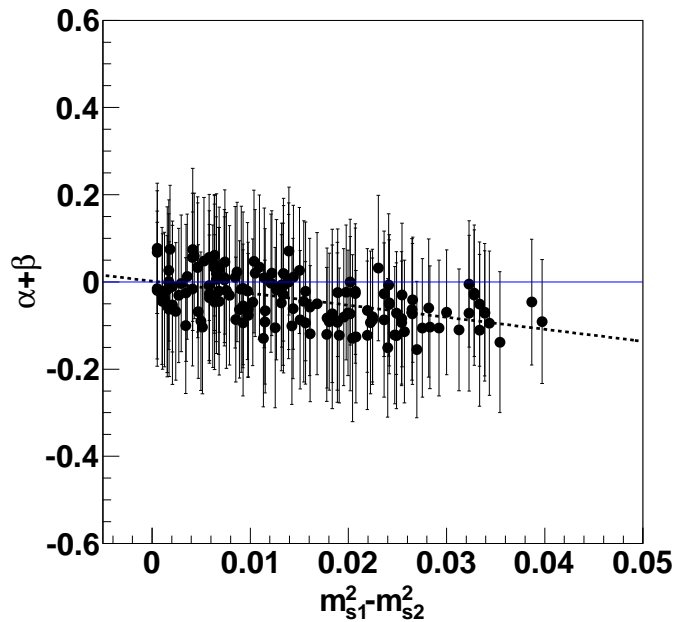}
\caption{\label{fig:fig_5}$\alpha(Z)+\beta(N)$ vs the difference (solid circles) 
in concentration for the two reaction systems for the case of Z=N=7. The dotted line is the 
result of a linear fit.}
\end{figure}

Standard isoscaling results have been derived under a general grand canonical approach~\cite{Ono03,Tsang01,Botvina02}. The Landau approach should be equivalent to it under certain conditions. Experimentally the b and c values have not been established because all isotopes identified in the present data have m $<$ 0.5 except for nucleons. In the case that b and c are of comparable magnitude to parameter a, which is assumed in the derivation of eq.(\ref{eq:hms}), we easily obtain:
\begin{equation}
  \label{eq:deltahms}
 \frac{\Delta H}{T}mA= a  \Delta m_s (N-Z) - \frac{1}{2}a (m^2_{s1}-m^2_{s2}) A =\alpha N+ \beta Z
 \end{equation}
 which introduces a volume term.  Equating similar terms we get:
 \begin{equation}
  \label{eq:alphabeta}
 \alpha=a \Delta m_s -\frac{1}{2}a (m^2_{s1}-m^2_{s2})
 \end{equation}
where $\Delta m_s=m_{s1}-m_{s2}$. It is straightforward to demonstrate the equivalence of the last equation to eq.(\ref{eq:alpha}) derived from the grand canonical approach.  In particular we get:
   \begin{equation}
  \label{eq:alpha+beta}
 \alpha+\beta=-a (m^2_{s1}-m^2_{s2})
 \end{equation}
 which shows that the two approaches are equivalent and that $m$ is an order parameter if $\alpha +\beta =0$ i.e. neglecting $O(m_s^2)$ terms in the external field.  In figure (\ref{fig:fig_5}) we plot $\alpha+\beta$ vs. $m^2_{s1}-m^2_{s2}$,  unfortunately the error bars are rather large but we can see a systematic deviation from zero as expected from eq.(\ref{eq:alpha+beta}) for large differences in concentration.
This indicates that, at the level of sensitivity so far  acheived with data of this  type the presence of higher order terms in m is difficult to quantify. Thus, within the error bars, $m$ could be considered an order parameter when relatively neutron  (or proton) rich
sources are considered.  In particular, phase transitions in finite systems could be studied using the same language of macroscopic systems i.e.,  'turning on and off' an external field~\cite{Bonasera08}.

\section *{ SUMMARY}

In conclusion, in this paper we have discussed scaling of ratios of 
yields from different colliding systems under similar physical 
control parameters, i.e. density and temperature. 
A careful and precise determination of isotopic yields is needed 
in order to see the features of the system near the phase transition. 
There is an order parameter, m, 
given by the difference in neutron and proton  concentrations  of the 
detected fragments which leads to an expected isoscaling relation, 
a direct consequence of the restored symmetry of the nuclear Hamiltonian
 when exchanging neutrons with protons.  The data suggest that the 
Coulomb field may not significantly violate such a symmetry. 
The existence of m-scaling might be a signature for near criticality 
of the fragmenting system. 
Other properties of the 'rich' nuclear Hamiltonian, such as pairing, 
appear to result in small violations of the scaling.  This is an 
interesting physical aspect which deserves further and deep 
investigation both theoretically
and experimentally.  Also, it would be interesting to search 
for m-scaling violations in heavily charged colliding systems 
such as U+U.  The absence of a violation in 
these cases would suggest that densities and deformations of the 
fragments are such that the effect of Coulomb is significantly reduced. 
Studies of the other extreme case of very exotic colliding systems 
would be also
be valuable to probe the effects of high 'external' field on the phase 
transition. The atomic nucleus constitutes a formidable laboratory to 
test our knowledge and understanding of phase transitions in a finite
system and offers a unique possibility for different quantum aspects 
similar to other bosons and fermion mixtures.


A major consideration in the interpretation of the results 
presented in this paper is the effect of the secondary decay
process. In the experiments excited fragments cool down to 
the ground state before they are detected. The reconstruction of the 
primary fragments from the experimentally observed IMFs and associated 
particles is not straightforward, since multiple excited primary 
fragments may be simultaneously produced in multifragmentation reactions 
making the unambiguous identification of the primary fragment distribution 
difficult. Indeed a major goal of the experiments from which the 
present isoscaling data are taken was to employ fragment-particle
correlation measurements to reconstruct the primary fragment 
distribution. The correlation data are still being analyzed~\cite{Wada05}.

\begin{acknowledgments}
This work is supported by the U.S. Department of Energy and the Robert 
A. Welch Foundation under grant A0330. One of us(Z. Chen) also thanks 
the \textquotedblleft100 
Persons Project" of the Chinese Academy of Sciences for the support. 
\end{acknowledgments}

\end{document}